\documentclass[aps,twocolumn]{revtex4}

\begin{document}
\begin{center}
{\Large \it Physical Review E 66 (2002) 056116}
\end{center}

\title{Pure Stationary States of Open Quantum Systems}

\author{\large Vasily E. Tarasov }

\affiliation{\it Skobeltsyn Institute of Nuclear Physics,
Moscow State University, Moscow 119992, Russia }

\email{tarasov@theory.sinp.msu.ru}

\begin{abstract}
Using Liouville space and superoperator formalism we consider
pure stationary states of open and dissipative quantum systems.
We discuss stationary states of open quantum systems, which
coincide with stationary states of closed quantum systems.
Open quantum systems with pure stationary states of linear oscillator
are suggested. We consider stationary states
for the Lindblad equation.
We discuss bifurcations of pure stationary states
for open quantum systems which are quantum analogs of
classical dynamical bifurcations.
\end{abstract}



\maketitle

\section{Introduction}

The open quantum systems are of strong theoretical interest.
As a rule, any microscopic system is always embedded in some
(macroscopic) environment and therefore it is never really
closed. Frequently, the relevant environment is in principle
unobservable or it is unknown \cite{Mash,T1}.
This would render the theory of open quantum systems
a fundamental generalization of quantum mechanics \cite{Dav,Prig}.

Classical open and dissipative systems can have regular or strange
attractors \cite{Hak,Lan}.
Regular attractors can be considered as a set of (stationary)
states for closed classical systems correspondent to open systems.
Quantization of evolution equations in phase space for dissipative
and open classical systems was suggested in Refs. \cite{Tarpla,Tarmsu}.
This quantization procedure allows one to derive quantum analogs
of open classical systems with regular attractors such as
nonlinear oscillator \cite{Tarpla,Tarpr2}.
In the papers \cite{Tarpla,Tarmsu,Tarpr2} were derived quantum analogs
of dissipative systems with strange attractors such as Lorenz-like
system, Rossler and Newton-Leipnik systems.
It is interesting to consider quantum analogs for
regular and strange attractors. The regular "quantum" attractors
can be considered as stationary states of open quantum systems.
The existence of stationary states for open quantum systems
is an interesting fact.
The condition given by Davies in Ref. \cite{Dav1}
defines the stationary state of open quantum system.
An example, where the stationary state is unique and approached by
all states for long times is considered by Lindblad \cite{Lind2} for
Brownian motion of quantum harmonic oscillator.
In Refs. \cite{Spohn,Spohn2,Spohn3} Spohn derives sufficient condition 
for the existence of a unique stationary state for the open quantum system
described by Lindblad equation \cite{Lind1}.
The stationary solution of the Wigner function evolution equation
for an open quantum system was discussed in Refs. \cite{AH,ISS}.
Quantum effects in the steady states of  the
dissipative map are considered in Ref. \cite{DC}.

In this paper we consider stationary pure states of some open
quantum systems. These open systems look like
closed quantum systems in the pure stationary states.
We consider the quantum analog of dynamical bifurcations
considered by Thompson and Lunn \cite{TL}
for classical dynamical systems.
In order to describe these systems, we consider Liouville-von Neumann
equation for density matrix evolution such that
this Liouville generator of the equation
is a function of some Hamiltonian operator.
Open quantum systems with pure stationary states
of linear harmonic oscillator are suggested.
We derive stationary states for quantum Markovian
master equation usually called the Lindblad equation.
The suggested approach allows one to use theory of
bifurcations for a wide class of quantum open systems.
We consider the example of bifurcation of pure
stationary states for open quantum systems.

In Section II, we introduce pure stationary states for closed
quantum systems and some mathematical background is considered.
In the Section III, we consider Liouville-von Neumann equation
for an open quantum system and pure stationary states for this equation.
In Section IV, simple examples stationary states for
open quantum systems are considered. In Section V,
we study some properties of quantum system to have
dynamical bifurcations and catastrophes.
In the Section VI, we suggest an example of quantum system with
fold catastrophe. Finally, a short conclusion is given in Section VII.
In Appendix, the mathematical backgroung (Liouville space, superoperators)
is considered.

\section{Pure stationary state}

In the general case, the time evolution of the quantum state $|\rho_t)$ 
can be described by the Liouville-von Neumann equation
\begin{equation}\label{LN}
\frac{d}{dt} |\rho_t)= \hat \Lambda  |\rho_t) , \end{equation}
where $\hat \Lambda$ is a Liouville superoperator on
Liouville space, $|\rho)$ is a density matrix operator
as an element of Liouville space.
For the concept of Liouville space
and superoperators see the Appendix and Refs. \cite{Cra}-\cite{kn2}.
For closed systems, Liouville superoperator has the form
\begin{equation} \label{hLo}
\hat \Lambda = -\frac{i}{\hbar}(\hat L_H- \hat R_H) \ \  or \ \
\hat \Lambda = \hat L^{-}_H , \end{equation}
where $H=H(q,p)$ is a Hamilton operator. If the Liouville superoperator
$\hat \Lambda$ cannot be represented in the form (\ref{hLo}), then
quantum system is called open, non-Hamiltonian or dissipative quantum
system \cite{kn1,kn2,Tartmf3}.
The stationary state is defined by the following condition: 
\begin{equation} \label{Lr0} \hat \Lambda |\rho_t)=0 . \end{equation}
For closed quantum systems (\ref{hLo}), this condition has the simple form:
\begin{equation}\label{s21}
\hat L_H|\rho_t)=\hat R_H |\rho_t) \ \ or \ \
\hat L^{-}_H |\rho_t)=0 \ . \end{equation}
In the general case, we can consider the Liouville superoperator as
a superoperator function \cite{Tarpla,Tarmsu,kn1}:
\[ \hat \Lambda=\Lambda(\hat L^{-}_X,\hat L^{+}_X) \ \ or \ \
\hat \Lambda=\Lambda(\hat L_X,\hat R_X) , \]
where $X$ is a set of linear operators. For example,
$X=\{q,p,H\}$ or $X=\{H_1,..,H_s\}$.
In the paper we use the special form of the superoperator
$\hat \Lambda$ such that
\[ \hat\Lambda=-\frac{i}{\hbar}(\hat L_H -\hat R_H) +
\sum^s_{k=1}  \hat F_k N_k(\hat L_H,\hat R_{H}) , \]
where $N_{k}(\hat L_H,\hat R_{H})$ are some superoperator functions
and  $\hat F^{k}$ is an arbitrary nonzero superoperator.

It is known that
a pure state $|\rho_{\Psi})$ is a stationary state of
a closed quantum system (\ref{LN}), (\ref{hLo}), if
the state $|\rho_{\Psi})$ is an eigenvector of the Liouville space
for superoperators $\hat L_H$ and $\hat R_H$:
\begin{equation} \label{LRH}
\hat L_H |\rho_{\Psi})=|\rho_{\Psi}) E , \ \
\hat R_H |\rho_{\Psi})=|\rho_{\Psi}) E . \end{equation}
Equivalently, the state $|\rho_{\Psi})$ is an eigenvector of
superoperators $L^{+}_H$ and $L^{-}_H$ such that
\[ \hat L^{+}_H |\rho_{\Psi})=|\rho_{\Psi}) E , \ \
\hat L^{-}_H |\rho_{\Psi})=|\rho_{\Psi}) \cdot 0=0 . \]
The energy variable $E$ can be defined by
\[ E=(I|\hat L_H |\rho_{\Psi})=(I|\hat R_H |\rho_{\Psi})
=(I|\hat L^{+}_H |\rho_{\Psi}) . \]

The superoperators $\hat L_{H}$ and $\hat R_{H}$ for
linear harmonic oscillator are
\begin{equation} \label{harm}
\hat L_H= \frac{1}{2m}\hat L^2_p+ \frac{m \omega^{2}}{2} \hat L^2_q ,
\ \
\hat R_H= \frac{1}{2m}\hat R^2_p+ \frac{m \omega^{2}}{2} \hat R^2_q .
\end{equation}
It is known that pure stationary states
$\rho_{\Psi_n}=\rho^{2}_{\Psi_n}$ of
linear harmonic oscillator (\ref{harm}) exists if the
variable $E$ is equal to
\begin{equation}\label{En}
E_{n}=\frac{1}{2}\hbar\omega(2n+1) \ . \end{equation}

\section{Pure stationary states of open systems}

Let us consider the Liouville-von Neumann equation (\ref{LN})
for the open quantum system defined of the form
\begin{equation}\label{h1}
\frac{d}{dt} |\rho_{t})=-\frac{i}{\hbar}(\hat L_H -\hat R_H)|\rho_{t}) +
\sum^s_{k=1}  \hat F_k N_k(\hat L_H,\hat R_{H}) |\rho_{t}) . \end{equation}
Here $\hat F^{k}$ is some superoperator
and $N_{k}(\hat L_H,\hat R_{H})$, where $k=1,...,s$,
are superoperator functions.

Let $|\rho_{\Psi})$ is a pure stationary state of the closed
quantum system defined by Hamilton operator $H$.
If Eqs. (\ref{LRH}) are satisfied, then the state
$|\rho_{\Psi})$
is a stationary state of the closed system
associated with the open system (\ref{h1}) and is defined by
\begin{equation}\label{h2} \frac{d}{dt} |\rho_{t})=-
\frac{i}{\hbar}(\hat L_H-\hat R_H) |\rho_{t}). \end{equation}
If the vector $|\rho_{\Psi})$ is an eigenvector of operators $\hat L_H$
and $\hat R_H$, then the Liouville-von Neumann equation (\ref{h1})
for the pure state $|\rho_{\Psi})$ has the form
\[ \frac{d}{dt} |\rho_{\Psi})=
\sum^s_{k=1} \ \hat F_{k} |\rho_{\Psi}) \ N_{k}(E,E) , \]
where the function $N_k(E,E)$ are defined by
\[ N_{k}(E,E)=(I|N_{k}(\hat L_H,\hat R_{H}) |\rho_{\Psi}) . \]
If all functions $N_{k}(E,E)$ are equal to zero
\begin{equation}\label{sc} N_{k}(E,E)=0 , \end{equation}
then the stationary state $|\rho_{\Psi})$ of the 
closed quantum system (\ref{h2})
is the stationary state of the open quantum system (\ref{h1}).

Note that functions $N_{k}(E,E)$ are eigenvalues and $|\rho_{\Psi})$ is
the eigenvector of superoperators $N_{k}(\hat L_{H},\hat R_{H})$, since
\[ N_{k}( \hat L_{H},\hat R_{H})|\rho_{\Psi})=|\rho_{\Psi})N_{k}(E,E) . \]
Therefore stationary states of the open quantum system (\ref{h1}) are
defined by zero eigenvalues of superoperators
$N_{k}(\hat L_{H},\hat R_{H})$.

\section{Open systems with oscillator stationary states}

In this section simple examples of open
quantum systems (\ref{h1}) are considered.

1) Let us consider the nonlinear oscillator with friction defined by
the equation
\begin{equation}\label{nlo1} \frac{d}{dt}  \rho_t=
-\frac{i}{\hbar}[ \tilde H,  \rho_t] - \frac{i}{2\hbar}\beta  [ q^2,
p^2 \rho_t+\rho_t p^2] , \end{equation}
where the operator $\tilde H$ is the Hamilton operator of the nonlinear
oscillator:
\[  \tilde H=\frac{ p^2}{2m}+
\frac{m \Omega^2  q^2}{2}+\frac{\gamma  q^4}{2} . \]
Equation (\ref{nlo1}) can be rewritten in the form
\[ \frac{d}{dt} |\rho_t)= \hat L^{-}_H |\rho_t)+ \]
\begin{equation}\label{nlo2}
+2m\beta \hat L^{-}_{q^2}
\Bigl( \frac{1}{2m}(\hat L^{+}_p)^2+\frac{\gamma}{2m\beta}
(\hat L^{+}_q)^2-\frac{\Delta}{4\beta} \hat L_I \Bigr)
|\rho_t) , \end{equation}
where $\Delta=\Omega^2-\omega^2$ and the superoperator
$\hat L^{-}_H$ is defined for the Hamilton operator $H$ of the 
linear harmonic oscillator by Eqs. (\ref{hLo}) and (\ref{harm}).
Equation (\ref{nlo2}) has the form (\ref{h1}), with

\[ N(\hat L_H,\hat R_H)=\frac{1}{2}(\hat L_H+\hat
R_H)- \frac{\Delta}{2\beta}\hat L_{I} , \quad
\hat F =2m\beta \hat L^{-}_{q^2} . \]
In this case the function $N(E,E)$ has the form
\[ N(E,E)=E-\frac{\Delta}{2 \beta}, \quad
 \]
Let $\gamma=\beta m^2\omega^2$.
The open quantum system (\ref{nlo1})
has one stationary state of the linear harmonic oscillator
with energy $E_n=(\hbar \omega /2)(2n+1)$, if
$\Delta=2\beta \hbar \omega (2n+1)$,
where $n$ is an integer non-negative number. This stationary state
is one of the stationary states of the linear harmonic
oscillator with the mass $m$ and frequency $\omega$. In this case
we can have the quantum analog of dynamical Hopf bifurcation
\cite{TL,MMC}.

2) Let us consider the open quantum system described
by the time evolution equation
\begin{equation}\label{h4} \frac{d}{dt} |\rho_{t})=
\hat L^{-}_H | \rho_{t})
+\hat L^{-}_q cos\Bigl(\frac{\pi}{\varepsilon_0} \hat L^{+}_{H} \Bigr)
|\rho_{t}) , \end{equation}
where the superoperator $\hat L^{-}_{H}$ is defined
by formulas (\ref{hLo}) and (\ref{harm}).
Equation (\ref{h4}) has the form (\ref{h1}) if
the superoperators $\hat F$ and $N(\hat L_H,\hat R_{H})$ are
defined by
\[ \hat F=-\frac{i}{\hbar}(\hat L_q-\hat R_q) , \]
\[ N(\hat L_H,\hat R_{H})=cos\Bigl(
\frac{\pi}{2\varepsilon_0} (\hat L_{H}+\hat R_{H})\Bigr) =\]
\begin{equation}\label{h4n} =\sum^{\infty}_{m=0}\frac{1}{(2m)!} \Bigl(
\frac{i\pi}{2\varepsilon_0}\Bigr)^{2m} (\hat L_{H}+\hat
R_{H})^{2m} . \end{equation}
The function $N(E,E)$ has the form
\[ N(E,E)=cos\Bigl( \frac{\pi E}{\varepsilon_0}\Bigr)=
\sum^{\infty}_{m=0}\frac{1}{(2m)!}
\Bigl( \frac{i\pi E}{\varepsilon_0}\Bigr)^{2m} . \]
The stationary state condition (\ref{sc})
has the solution
\[ E=\frac{\varepsilon_0}{2} (2n+1) , \]
where $n$ is an integer number.
If parameter $\varepsilon_{0}$ is equal to $\hbar \omega$, then
quantum system (\ref{h4}) and (\ref{h4n}) has pure stationary states
of the linear harmonic oscillator with the energy (\ref{En}).
As the result, stationary states of the open quantum system (\ref{h4})
coincide with pure stationary states of linear harmonic oscillator.
If the parameter $\varepsilon_{0}$ is equal to $\hbar
\omega(2m+1)$, then quantum system (\ref{h4}) and (\ref{h4n}) have
stationary states of the linear harmonic oscillator
with $n(k,m)=2km+k+m$ and
\[ E_{n(k,m)}=\frac{\hbar \omega}{2}(2k+1)(2m+1) . \]

3) Let us consider the superoperator function
$N_{k}(\hat L_{H},\hat R_{H})$ in the form
\[ N_{k}(\hat L_{H},\hat R_{H})= 
\frac{1}{2\hbar} \sum_{n,m} v_{kn} v^{*}_{km}
( 2\hat L^{n}_{H} \hat R^{m}_{H}-
\hat L^{n+m}_{H}-\hat R^{n+m}_{H} ) , \]
and all superoperators $\hat F_{k}$
are equal to $\hat L_{I}$. In this case,
the Liouville-von Neumann equation (\ref{h1}) can be represented by
the Lindblad equation \cite{Lind,AL,kn1}:
\[ \frac{d}{dt}|\rho_t)=-\frac{i}{\hbar}(\hat L_{H}-\hat R_{H}) |\rho_t)+\]
\begin{equation} \label{Lin} +\frac{1}{2 \hbar}
\sum_{j}\Bigl( 2\hat L_{V_{k}} \hat R_{V^{\dagger}_{k}}-
\hat L_{V^{ }_{k}} \hat L_{V^{\dagger}_{k}}-
\hat R_{V^{\dagger}_{k}} \hat R_{V^{ }_{k}} \Bigr)|\rho_t). \end{equation}
with linear operators $V_{k}$ defined by
\begin{equation} \label{Lin2} V_{k}=\sum_{n}v_{kn}H^{n} , \ \
V^{\dagger}_{k}=\sum_{m}v^{*}_{km}H^{m} . \end{equation}
If $|\rho_{\Psi})$ is a pure stationary state (\ref{LRH}), then
all functions $N_{k}(E,E)$ are equal to zero
and this state $|\rho_{\Psi})$
is a stationary state of the open quantum system (\ref{Lin}).

If the Hamilton operator $H$ is defined by
\begin{equation}
\label{nlo} H=\frac{1}{2m}p^{2}+\frac{m\omega^{2}}{2}q^{2}+
\frac{\lambda}{2} (qp+pq), 
\end{equation}
then we have some generalization of the quantum model for the Brownian motion
of a harmonic oscillator considered in \cite{Lind2}.
Note that in the model \cite{Lind2} operators $V_{k}$ are linear
$V_{k}=a_{k}p+b_{k}q$, but in our generalization (\ref{Lin}) and (\ref{Lin2})
these operators are nonlinear. For example, we can use
\[ V_{k}=a_{k}p+b_{k}q+c_kp^2+d_kq^2+e_k(qp+pq). \]

The case $c_k=d_k=e_k=0$ is considered in Ref. \cite{Lind2}.
Let real parameters $\alpha$ and $\beta$ exist and
\[ b_k=\alpha a_k, \quad c_k=\beta a_k, \quad 
d_k=m^2 \omega^2 \beta a_k, \quad e_k=m \lambda \beta a_k .  \]
In this case, the pure stationary states of the linear 
oscillator (\ref{nlo}) exist if $\omega > \lambda$ and 
the variable $E$ is equal to
\[ E_n=\frac{1}{2} \hbar \omega (2n+1)\sqrt{1-\lambda^2/\omega^2} .\]

\section{Dynamical bifurcations and catastrophes}

Let us consider a special case of open quantum systems (\ref{h1})
such that the vector function
\[ N_k(E,E)=(I|N_k(\hat L_H,\hat R_H) |\rho) , \]
be a potential function
and the Hamilton operator $H$ can be represented in the form
\[  H=\sum^s_{k=1}  H_k . \]
In this case we have a function $V(E)$ called potential, 
such that the following conditions are satisfied:
\[ \frac{\partial V(E)}{\partial E_{k}}=N_{k}(E,E) . \]
where $E_k=(I| \hat L_{H_k}|\rho)=(I| \hat R_{H_k}|\rho)$.
If potential $V(E)$ exists, then
the stationary state condition (\ref{sc}) for the 
open quantum system (\ref{h1}) is defined by critical points of
the potential $V(E)$. If the system has one variable $E$, then the
function $N(E,E)$ is always a potential function.
In general, the vector function $N_{k}(E,E)$ is potential, if
\[ \frac{\partial N_k(E,E)}{\partial E_l}= \frac{\partial
N_l(E,E)}{\partial E_k} . \]
Stationary states of the open quantum system (\ref{h1})
with potential vector function $N_{k}(E,E)$ is depend by critical
points of the potential $V(E)$. It allows one to use the theory of
bifurcations and catastrophes for the parametric set of functions
$V(E)$. Note that a bifurcation in a vector space of variables
$E=\{E_k|k=1,...,s \}$ is a bifurcation in the vector space of
eigenvalues of the Hamilton operator $H_k$.

For the polynomial superoperator function
$N_{k}(\hat L_H,\hat R_{H})$ we have
\[ N_{k}(\hat L_H,\hat R_{H})  =
\sum^{N}_{n=0} \sum^{n}_{m=0} a^{(k)}_{n,m}
\hat L^{m}_{H} \hat R^{n-m}_{H} . \]
In general case, $m$ and $n$ are multi-indices. The function
$N_{k}(E,E)$ is a polynomial
\[ N_{k}(E,E)=\sum^{N}_{n=0} \alpha^{(k)}_{n} E^{n} , \]
where the coefficients $\alpha^{(k)}_{n}$ are defined by
\[ \alpha^{(k)}_{n}=\sum^{n}_{m=0} a^{(k)}_{n,m} . \]

We can define the variables $x_l=E_l-a_l$ $(l=1,...,s)$, such that
functions $N_k(E,E)=N_k(x+a,x+a)$ have no the terms $x^{n-1}_l$.
\[ N_k(x+a,x+a)=\sum^{N}_{n=0} \alpha^{(k)}_{n} (x+a^{(k)})^{n}= \]
\[=\sum^{N}_{n=0} \sum^{n}_{m=0} \alpha^{(k)}_{n}
\frac{n!}{m!(n-m)!}x^{m}(a^{(k)})^{n-m} . \]
If the coefficient of the term $x^{n_l-1}_l$ is equal to zero
\[ \alpha^{(k)}_{n_l}\frac{n_l!}{(n_l-1)!}a^{(k)}_l +\alpha^{(k)}_{n_l-1}=
\alpha^{(k)}_{n_l} n_l a^{(k)}_l +\alpha^{(k)}_{n_l-1}=0  , \]
then we have the following coefficients
\[ a^{(k)}_l=-\frac{\alpha^{(k)}_{n_l-1}}{n_l \alpha^{(k)}_{n_l}}  . \]

If we change parameters $\alpha^{(k)}_n$, then an open quantum
system can have pure stationary states of the system.
For example, the bifurcation with the birth of
linear oscillator pure stationary state is a quantum analog of
dynamical Hopf bifurcation \cite{TL,MMC} for classical
dynamical system.

Let a vector space of energy variables $E$ be a
one-dimensional space.
If the function $N(E,E)$ is equal to
\[ N(E,E)=\pm \alpha_{n}  E^n+ \sum^{n-1}_{j=1}
\alpha_{j} E^{j} \ \ \ n \ge 2 , \]
then the potential $V(x)$ is defined by the following equation: 
\[ V(x)=\pm x^{n+1}+ \sum^{n-1}_{j=1}a_{j}x^{j} \ \ \ n \ge 2 , \]
and we have catastrophe of type $A_{\pm n}$.

If we have $s$ variables $E_l$, where $l=1,2,...,s$, then quantum
analogs of elementary catastrophes $A_{\pm n}$, $D_{\pm n}$,
$E_{\pm 6}$, $E_7$ and $E_8$ can be realized
for open quantum systems. Let us write the
full list of potentials $V(x)$, which leads to
elementary catastrophes (zero modal) defined by $V(x)=V_0(x)+Q(x)$, where
\[ A_{\pm n}: \ V_0(x)=\pm x^{n+1}_1+\sum^{n-1}_{j=1}a_{j}x^{j}_1 \ \
n \ge 2 , \]
\[ D_{\pm n}: \ V_0(x)=x^{2}_{1}x_{2} \pm x^{n-1}_2+
\sum^{n-3}_{j=1}a_{j}x^{j}_{2}+\sum^{n-1}_{j=n-2}x^{j-(n-3)}_{1}
, \]
\[ E_{\pm 6}: \ V_0(x)=(x^{3}_{1} \pm x^{4}_{2})+
\sum^{2}_{j=1}a_{j}x^{j}_{2}+\sum^{5}_{j=3}a_{j}x_{1}x^{j-3}_2 , \]
\[ E_{7}: \ V_0(x)=x^{3}_{1}+x_{1}x^{3}_{2}+
\sum^{4}_{j=1}a_{j}x^{j}_{2}+\sum^{6}_{j=5}a_{j}x_{1}x^{j-5}_2 , \]
\[ E_{8}: \ V_0(x)=x^{3}_{1}+x^{5}_{2}+
\sum^{3}_{j=1}a_{j}x^{j}_{2}+\sum^{7}_{j=4}a_{j}x_{1}x^{j-4}_2 . \] 
Here $Q(x)$ is the nondegenerate quadratic form with variables
$x_{2}$, $x_{3}$, ..., $x_{s}$ for $A_{\pm n}$ and parameters
$x_{3}$, ..., $x_{s}$ for other cases.

\section{Fold catastrophe}

In this section, we suggest an example of the open quantum system
with catastrophe $A_{2}$ called fold.

Let us consider the Liouville-von Neumann equation (\ref{h1})
for a nonlinear quantum oscillator with friction,
where multiplication superoperators $\hat L_H$ and $\hat R_{H}$
are defined by Eq. (\ref{harm}) and superoperators $\hat F$
and $N(\hat L_{H},\hat R_{H})$ are given by the following equations: 
\begin{equation} \label{F} \hat F=-2 \hat L^{-}_q \hat L^{+}_p  \ ,
\end{equation}
\begin{equation}
\label{h3} N(\hat L_{H},\hat R_{H})=\alpha_0 \hat L^{+}_I+\alpha_1
\hat L^{+}_{H}+ \alpha_2 (\hat L^{+}_{H})^2 , \end{equation}
In this case, the function $N(E,E)$ is equal to
\[ N(E,E)=\alpha_0+\alpha_1 E+\alpha_2 E^2 . \]

A pure stationary state $|\rho_{\Psi})$ of the linear harmonic oscillator
is a stationary state of the open quantum system (\ref{h3}), if
$N(E,E)=0$.
Let us define the new real variable $x$ and parameter $\lambda$ by
the following equation
\[ x=E+\frac{\alpha_{1}}{2\alpha_{2}} , \ \ \lambda=
\frac{4\alpha_{0}\alpha_{2}-\alpha^{2}_{1}}{4\alpha^{2}_{2}} . \]
Then we have the stationary condition $N(E,E)=0$ in the form
$x^{2}-\lambda=0$.
If $\lambda \le 0$, then  the open quantum system has no stationary states.
If $\lambda >0$, then we have pure stationary states for a discrete set
of parameter values $\lambda$.
If the parameters $\alpha_{1},\alpha_{2}$ and $\lambda$ satisfy the
following conditions
\[ -\frac{\alpha_{1}}{2\alpha_{2}}=\hbar \omega(n+\frac{1}{2}+\frac{m}{2}), \ \
\lambda=\hbar^{2} \omega^{2} \frac{m^{2}}{4} , \]
where $n$ and $m$ are non-negative integer numbers, then the open
quantum system (\ref{F}) and (\ref{h3}) has two pure stationary state 
of the linear harmonic oscillator.
The energies of these states are equal to
\[ E_n=\hbar \omega(n+\frac{1}{2}) \ ,
\quad E_{n+m}=\hbar \omega(n+m+\frac{1}{2}) . \]

\section{Conclusion}

Open quantum systems can have pure stationary states. Stationary states
of open quantum systems can coincide with pure stationary states of
closed (Hamiltonian) systems. As an example, we suggest open
quantum systems with pure stationary states of linear oscillator.
Note that using Eq. (\ref{h1}), it is easy to get open (dissipative)
quantum systems with stationary states of hydrogen atom. For a
special case of open systems, we can use usual bifurcation and
catastrophe theory. It is easy to derive quantum analogs of
classical dynamical bifurcations.
Physical instances with bifurcation behavior can be realized
in quantum optics \cite{Hak2} and deep inelastic collisions \cite{SS}.

Open quantum systems with two stationary states
can be considered as qubit. It allows one to consider
open n-qubit quantum system described by Eq. (\ref{h1}) 
as a quantum computer with pure states.
In general, we can consider open n-qubit quantum systems
as quantum computer with mixed states \cite{TarJPA,Tarpr3}.
A mixed state (operator of density matrix) of n two-level quantum
systems (open or closed n-qubit system) is an element
of $4^{n}$-dimensional operator Hilbert space (Liouville space).
It allows one to use quantum computer
model with four-valued logic \cite{TarJPA,Tarpr3}. The quantum gates
of this model are real, completely positive, trace-preserving
superoperators that act on mixed state.
Bifurcations of pure quantum states can used for quantum gate
control.


\section*{Acknowledgement}

This work was partially supported by the RFBR Grant No. 02-02-16444.

\section*{Appendix}

For the concept of Liouville space and superoperators see
Refs. \cite{Cra}-\cite{kn2}

\subsection*{Liouville space}

The space of linear operators acting on a Hilbert space ${\cal H}$
is a complex linear space $\overline {\cal H}$.
We denote an element $A$ of $\overline{\cal H}$ by a ket-vector $|A)$.
The inner product of two elements $|A)$ and
$|B)$ of $\overline{\cal H}$ is defined as
$(A|B)=Tr(A^{\dagger} B)$.
The norm $\|A\|=\sqrt{(A|A)}$ is the Hilbert-Schmidt norm of
operator $A$. A new Hilbert space $\overline{\cal H}$ with the inner
product  is called Liouville space attached to ${\cal
H}$ or the associated Hilbert space, or Hilbert-Schmidt space
\cite{Cra}-\cite{kn2}.

Let $\{|x>\}$ be an orthonormal basis of ${\cal H}$:
\[ <x|x'>=\delta(x-x') \ , \quad \int dx |x><x|=I . \]
Then $|x,x')=||x><x'|)$
is an orthonormal basis of the Liouville space
$\overline{\cal H}$:
\[ (x,x'|y,y')=\delta(x-x')\delta(y-y'), \]
\begin{equation} \label{onb}
\int dx \int dx' |x,x')(x,x'|=\hat I . \end{equation}
For an arbitrary element $|A)$ of $\overline{\cal H}$ we have
\begin{equation} \label{|A)}
|A)=\int dx \int dx' |x,x')(x,x'|A) \end{equation}
where
\[ (x,x'|A)=Tr( (|x><x'|)^{\dagger}A )= \]
\[ =Tr( |x'><x| A )=<x|A|x'>=A(x,x') , \]
is a kernel of the operator $A$.
An operator $\rho$ of density matrix
($Tr\rho=1$, $\rho^\dagger=\rho$, $\rho \ge 0$) can be considered
as an element $|\rho)$ of the Liouville space $\overline{\cal H}$.
Using (\ref{|A)}), we get
\begin{equation} \label{|rho)} |\rho)=
\int dx \int dx' |x,x')(x,x'|\rho) \ , \end{equation}
where the trace is represented by
\[ (I|\rho)= Tr\rho=\int dx \ (x,x|\rho)=1. \]

\subsection*{Superoperators}

Operators that act on $\overline{\cal H}$ are called superoperators
and we denote them, in general, by the hat.

For an arbitrary superoperator $\hat \Lambda$ on
$\overline{\cal H}$, which is defined by
\[ \hat \Lambda|A)=|\hat \Lambda(A) ), \]
we have
\[ (x,x'|\hat \Lambda|A)=\int dy \int dy'
(x,x'|\hat \Lambda|y,y') (y,y'|A)=\]
\[ =\int dy \int dy'
\Lambda(x,x',y,y') A(y,y') , \]
where $\Lambda(x,x',y,y')$ is a kernel of the superoperator $\hat \Lambda$.

Let $A$ be a linear operator in the Hilbert space ${\cal H}$.
We can define the multiplication superoperators
$\hat L_{A}$ and $\hat R_{A}$ by the following equations
\[ \hat L_{A}|B)=|AB) \ , \quad \hat R_{A}|B)=|BA). \]

The superoperator kernels can be easy derived. For example,
in the basis $|x,x')$ we have
\[ (x,x'|\hat L_{A}|B)=\int dy \int dy'
(x,x'|\hat L_{A}|y,y')(y,y'|B)=\]
\[ =\int dy \int dy' L_{A}(x,x',y,y') B(y,y'). \]
Using
\[ (x,x'|AB)=<x|AB|x'>=\]
\[ =\int dy \int dy' <x|A|y><y|B|y'><y'|x'> , \]
we get kernel of the left multiplication superoperator
\[ (\hat L_{A})(x,x',y,y') =<x|A|y><x'|y'>=A(x,y)\delta(x'-y'). \]


Left superoperators $\hat L^{\pm}_{A}$ are defined
as Lie and Jordan multiplication by
\[ \hat L^{-}_A B=\frac{1}{i\hbar}(AB-BA) , \ \
\hat L^{+}_AB=\frac{1}{2}(AB+BA) . \]
The left superoperators $\hat L^{\pm}_A$ and
right superoperators $\hat R^{\pm}_A$ are connected by
\[ \hat L^{-}_A=-\hat R^{-}_A , \ \ \hat L^{+}_A=\hat R^{+}_A . \]
An algebra of the superoperators $\hat L^{\pm}_{A}$
is defined \cite{Tarmsu} by the following relations \\
1) Lie relations
\[ \hat L^{-}_{A \cdot B}=\hat L^{-}_{A} \hat L^{-}_{B}-
\hat L^{-}_{B} \hat L^{-}_{A} . \]
2) Jordan relations
\[ \hat L^{+}_{(A \circ B)\circ C}+\hat L^{+}_{B}\hat L^{+}_{C}\hat L^{+}_{A}
+\hat L^{+}_{A}\hat L^{+}_{C}\hat L^{+}_{B}= \]
\[ =\hat L^{+}_{A\circ B}\hat L^{+}_{C}+
\hat L^{+}_{B\circ C}\hat L^{+}_{A}+\hat L^{+}_{A\circ C}\hat L^{+}_{B} , \]
\[ \hat L^{+}_{(A\circ B)\circ C}+\hat L^{+}_{B}\hat L^{+}_{C}\hat L^{+}_{A}+
\hat L^{+}_{A}\hat L^{+}_{C}\hat L^{+}_{B}= \]
\[ =\hat L^{+}_{C}\hat L^{+}_{A\circ B}+\hat L^{+}_{B}\hat L^{+}_{A\circ C}+
\hat L^{+}_{A}\hat L^{+}_{B\circ C} , \]
\[ \hat L^{+}_{C}\hat L^{+}_{A\circ B}+\hat L^{+}_{B}\hat L^{+}_{A\circ C}+
\hat L^{+}_{A}\hat L^{+}_{B\circ C}= \]
\[ =\hat L^{+}_{A\circ B}\hat L^{+}_{C}+
\hat L^{+}_{B\circ C}\hat L^{+}_{A}+\hat L^{+}_{A\circ C}\hat L^{+}_{B} . \]
3) Mixed relations
\[ \hat L^{+}_{A \cdot B}=
\hat L^{-}_{A}\hat L^{+}_{B}-\hat L^{+}_{B}\hat L^{-}_{A} , \]
\[ \hat L^{-}_{A \circ B}=\hat L^{+}_{A}\hat L^{-}_{B}+
\hat L^{+}_{B}\hat L^{-}_{A} , \]
\[ \hat L^{+}_{A \circ B}=\hat L^{+}_{A}\hat L^{+}_{B}-\frac{\hbar^{2}}{4}
\hat L^{-}_{B}\hat L^{-}_{A} , \]
\[ \hat L^{+}_{B}\hat L^{+}_{A}-\hat L^{+}_{A}\hat L^{+}_{B}=
-\frac{\hbar^{2}}{4} \hat L^{-}_{A \cdot B} , \]
where
\[ A \cdot B=\frac{1}{i \hbar}(AB-BA), \ \
 A \circ B=\frac{1}{2}(AB+BA) . \]



\end{document}